\documentclass[twocolumn]{aastex631}

\newcommand\astrowisp{\texttt{AstroWISP}}
\newcommand\autowisp{\texttt{AutoWISP}}
\newcommand\fitstarshape{\texttt{FitStarShape}}
\newcommand\subpixphot{\texttt{SubPixPhot}}

\renewcommand{\noprint}[1]{}

\usepackage{import}
\usepackage{relsize}

\graphicspath{ {figures/} }

\begin{document}

\title{Tools for High-Precision Photometry from Wide-Field Color Images}

\author[0000-0003-4464-1371]{Kaloyan Penev}
\author[0000-0001-9436-6027]{Angel Romero}
\author[0000-0002-9490-2093] {S. Javad Jafarzadeh}
\affiliation{
    University of Texas at Dallas,
    800 W. Campbell Road,
    Richardson, TX 75080-3021, USA
}

\author[0000-0002-1097-9908] {Olivier Guyon}
\affiliation{
    Subaru Telescope, National Astronomical Observatory of Japan,
    650 N Aohoku Pl,
    Hilo, Hawaii 96720, USA
}
\affiliation{
    Steward Observatory, The University of Arizona,
    933 N Cherry Ave,
    Tucson AZ 85721, USA
}
\affiliation{
    Wyant College of Optical Sciences, The University of Arizona,
    1630 E University Blvd,
    Tucson, AZ 85721, USA
}

\author[0000-0002-2931-7605]{Wilfred Gee}
\author[0000-0001-5320-8049]{Preethi Krishnamoorthy}
\affiliation{
    Subaru Telescope, National Astronomical Observatory of Japan,
    650 N Aohoku Pl,
    Hilo, Hawaii 96720, USA
}


\begin{abstract}
We present \astrowisp: a collection of image processing tools for source
extraction, background determination, point-spread function/pixel-response
function fitting, and aperture photometry. \astrowisp{} is particularly
well-suited for working with detectors featuring a Bayer mask (an array of
microfilters applied to each detector pixel to allow color photography), such as
consumer DSLR cameras. Such detectors pose significant challenges for existing
tools while offering a much cheaper alternative to specialized devices. As a
result, consumer digital single-lens reflex (DSLR) cameras with Bayer masks are
often underutilized for precision photometry. \astrowisp{} addresses this
limitation in an effort to democratize precision photometry and support broader
community participation in research. We demonstrate that our tools produce
high-precision photometry from such images, enabling the use of such devices for
detecting exoplanet transits.  We package our tools for all major operating
systems to ensure accessibility for amateur astronomers.

\end{abstract}



\section{Introduction}%
\label{sec:intro}

Brightness measurements are the observing technique  most accessible to citizen
scientists. The venerable American Association of Variable Star Observers has
been promoting citizen scientist photometry for more than a century, and in the
area of exoplanets, a number of projects are working to harness the power of
photometric observations by citizen scientists, like the JPL-lead Exoplanet
Watch (EW) \citep{Zellem_et_al_19, Zellem_et_al_20}, PANOPTES
\citep{Guyon_et_al_14}, and the Unistellar network \citep{Marchis_et_al_20,
Esposito_et_al_21}. These and numerous other examples of citizen scientists
providing observational data demonstrate that amateur astronomers are eager to
contribute their skills, time, and equipment to advance science.

Citizen scientists are particularly well suited to pursue the detection and
confirmation of planets with long orbital periods and deep transits, orbiting
very bright host stars. A search of the NASA exoplanet archive shows fewer than
20 planets with $V<11\,mag$ (TESS magnitude $<10\,mag$), transit depth $>0.2\%$,
and orbital period longer than 15 days. However, bright hosts and deep transits
make these planets ideal candidates for follow-up characterization using radial
velocity (RV) measurements, transit and secondary eclipse spectroscopy,
phase-curve analysis, Rossiter-McLaughlin observations, and a host of other
observational techniques that probe a wide range of the physical processes in
the atmosphere, or the bulk of the planet.

Studying longer period planets will dramatically enhance our ability to
interpret a wide range of follow-up observations and improve our understanding
of planetary physics. For example, the anomalously large sizes of short-period
giant planets have been difficult to explain, and irradiation from the star
appears to be strongly implicated \citep[see][]{Demory_Seager_11}.  Irradiation
also affects the thermal structure and chemistry of the outer layers of planets,
influencing the observed transmission spectra or multi-band transit measurements
\citep[see[]{Kawashima_Ikoma_19}. Tidal and magnetic interactions also
confound the interpretation of observations \citep[see][]{Strugarek_18}. These
effects will be several orders of magnitude weaker for planets with periods of
order weeks or months relative to the current best targets for detailed planet
or atmospheric characterization. This will allow the signatures of star-planet
interactions to be isolated from processes internal to the planet, enabling
in-depth investigations of both.

Citizen science projects reside in a unique niche of very large data volume, yet
limited data quality. For example, calibration data is often not collected, the
instruments are much less sensitive and stable compared to professional ones,
data collection conditions are subject to light pollution or high weather
intermittency, the detectors (often off-the-shelf color digital single-lens
reflex (DSLR) cameras) have higher noise and lower sensitivity, and crucially a
staggered array of microfilters residing on top of pixels to allow color
photometry. All of this necessitates flexible image processing and denoising
techniques, capable of cleaning defects in the data as much as possible.

Our research group is embarking on a project to enable high-precision photometry
by citizen scientists. In this first article, we present a basic set of tools
allowing the use of color detectors like DSLR cameras. Thanks to mass production
and built-in firmware and software, such detectors have $\sim 10\times$ lower
cost per etendue and are dramatically simpler to control and operate compared to
traditional CCD detectors. These features make it possible for citizen
scientists to assemble their own observing platforms and carry out photometric
surveys.

\begin{figure}[ht!]
    \begin{center}
        \includegraphics[width=0.4\textwidth]{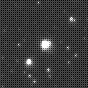}\\
        \vspace{3mm}
        \includegraphics[width=0.4\textwidth]{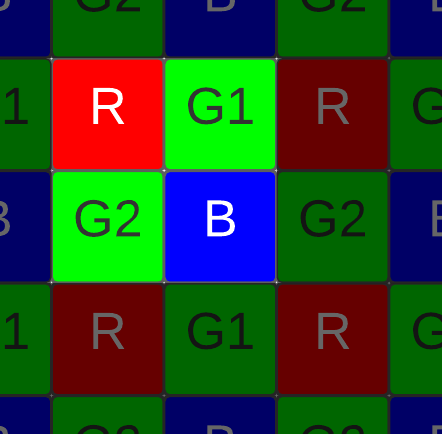}
    \end{center}
    \caption{%
        Top: $88\times88$ pixel portion, centered on WASP-33, of a
        $5208\times3476$ pixel image taken with a DSLR camera. The pattern is
        due to the Bayer mask. Bottom: arrangement of pixels sensitive to
        different colors in super-pixels (highlighted area) in a DSLR detector.%
        \label{fig:color_image}
    }
\end{figure}

However, observations made with color cameras and typically low-cost telescopes
pose a set of unique difficulties for image processing. The most important
challenge to extracting high-precision photometry from color cameras compared to
monochrome detectors is that they have a separate microfilter on top of each
pixel with the three colors alternating on neighboring pixels, a.k.a.\ a Bayer
mask (see Fig.~\ref{fig:color_image}). This poses grave challenges to existing
photometry software, because an image of a single color channel only comes from
1/4 of the image area (Bayer masks are organized in four channels: red, blue,
and two green ones), and how a star lines up with that sub-set of pixels
dramatically changes the response, especially if the image of the star is sharp
(i.e.\ the vast majority of the light falls on only a few pixels). Secondly,
light hitting the pixels of one channel sometimes changes the response of
neighboring pixels (belonging to different channels).  For example, saturating a
pixel can cause electrons to bleed to its neighbours, or through the so called
brighter-fatter effect (see \citep{Antilogus_et_al_14}).  Thirdly, the pixels
typical of our target detectors can hold fewer electrons and their readout
electronics have lower resolution. For example, professional cameras typically
have 16 bit amplifiers (sometimes more), while most DSLRs are 14 bit or fewer.
The result is a narrower dynamic range over which precise photometry is even
theoretically possible. Fourthly, for many CCDs the signal that would be
registered by a saturated pixel leaks to its neighbors.  This makes it possible
to extend accurate photometry to moderately saturated stars by tracking and
including the leaked flux into the measurement.  DSLR detectors (and some CCDs)
on the other hand have an ``antiblooming'' feature, which removes this excess
charge. Even worse, this ``feature'' starts removing charge even before a pixel
is truly saturated, causing nonlinearity, and further narrowing the dynamic
range.

Many tools already exist for extracting photometry from digital images (in fact
too many to list). However, none of these tools were designed with color
detectors in mind. The best photometry from color images to date was published
by \citet{Guyon_Martinache_12}, \citet{Zhang_et_al_15}, and
\citet{Gee_et_al_21}, all requiring the development of custom photometry
techniques, in lieu of all existing tools.

In this article, we present a collection of tools we call \astrowisp{}:
\textbf{Astro}nomical \textbf{W}ide-field \textbf{I}mages \textbf{S}tellar
\textbf{P}hotometry \citep{astrowisp_1.0}, which significantly improves upon the
previous efforts mentioned above.

The tasks performed by our tools are:
\begin{itemize}
    \item Source extraction, to be used for matching to an existing catalog to
        derive a transformation from sky coordinates to image coordinates.
    \item Background extraction.
    \item Point-spread function (PSF)/pixel-response function (PRF) fitting
        photometry.
    \item Aperture photometry.
\end{itemize}
For each of these tasks, \astrowisp{} properly propagates the uncertainty in
each image pixel to the final measurement reported.

\section{The \astrowisp{} Suite}
A core theme of \astrowisp{} is accounting for the fact that detector pixels are
not densely packed squares, uniformly sensitive to light over their entire
surface. As a result, how the PSF of a star lines up with the pixel boundaries
affects what fraction of that star's light is registered by the detector.
\astrowisp{} handles the effect of this nonuniform subpixel sensitivity without
any approximation using a highly efficient implementation of exact analytical
expressions for the integrals of arbitrary polynomials over all possible
overlaps between circles and rectangles. Accounting for nonuniformly sensitive
pixels is especially crucial for color images. Taking the pixels of only one
color channel, and treating it as an image with densely packed pixels is
equivalent to an image where each pixel is sensitive to light over only 1/4 of
its area (the other 3/4 belong to other channels). In \astrowisp{}, photometry
is extracted using aperture photometry and PSF fitting. This is performed
optimally, correcting for the Bayer mask, PSF, source positions, and other
relevant information (e.g., information from an external catalog). All tools are
implemented in C/C++ for efficiency, but we provide a convenient Python
interface for use in a data processing pipeline (currently under development).
An important requirement imposed by targeting citizen scientists is to make any
software available for all three major operating systems: Windows, MacOS, and
Linux. \astrowisp{} is available as a Python package with pre-built wheels for
multiple versions of all three operating systems, and multiple Python versions.
The software is specifically designed to allow us to easily add support as new
versions of Python or operating systems become available.

The data processing tasks implemented in \astrowisp{} are not sufficient to
provide full service processing of images to high-quality lightcurves.  The
envisioned application of \astrowisp{} is for it to be integrated into such full
processing pipelines that are specifically tuned to specific projects or
applications. Since the vast majority of astronomical software is in Python, we
provide a Python interface to the functionality of \astrowisp{}, described in
the package documentation \footnote{\url{https://kpenev.github.io/AstroWISP/}}.
We provide such full-service processing, including things like image calibration
(to remove bias, dark current, flat-field variability), astrometry, and various
forms of detrending and systematics removal, as a separate Python package ---
\autowisp{} --- which is described in a companion paper \citep{Romero_et_al_25},
available as a Python package
\footnote{\url{https://pypi.org/project/autowisp/}}.

One reason for splitting the software in two packages is to support integration
into data processing pipelines by other groups. Another reason is the
requirement to support multiple operating systems. Including compiled code in a
Python package in a user-friendly manner requires distributing separate Python
wheels (precompiled versions) for each version of each operating system.  This
makes generating new releases of such packages labor intensive. We thus elect to
separate the compiled part of the software, namely \astrowisp, into a separate
package that is expected to remain much more stable over time from Python-only
components (\autowisp) that are expected to evolve much more.

The tools described here are not only useful for citizen scientists or just
color detectors. The theme of accounting for nonuniformly sensitive pixels,
dealing with under or critically sampled point spread function images, and
obsessively tracking the error budget throughout the photometry process can
benefit professional surveys as well, including ones using monochrome detectors.

An outline of the rest of the paper is as follows: Section~\ref{sec:psf_fitting}
describes the background extraction and PSF/PRF fitting tool,
Section~\ref{sec:aperture_photometry} describes the aperture photometry tool,
Section~\ref{sec:python_wrapper} describes the Python interface to these tools,
Section~\ref{sec:packaging} describes our scheme for packaging and continuous
integration, Section~\ref{sec:demo} presents the performance of \astrowisp{} in
processing real-world data from color cameras, and Section~\ref{sec:discussion}
offers a summary and concluding remarks.

\subsection{Point-Spread Function/Pixel-Response Function Fitting}%
\label{sec:psf_fitting}

\astrowisp{} follows the strategy of professional transiting planet surveys of
deriving a transformation between an existing much higher-resolution catalog
(e.g. Gaia) and the stars extracted from the image (a.k.a astrometry). This
transformation is then used to project the catalog positions on the frame. This
results in source positions that are much more precise and accurate than what
source extraction provides directly from the image, because the transformation
is a smooth function with far fewer parameters than the number of stars used to
derive it. Furthermore, using an external, much deeper and higher-resolution
catalog provides valuable information on blended sources and source properties,
which are then used to improve photometry extraction.

Finding an astrometric solution requires a tool to automatically extract
stars from images. While there is no shortage of such tools, our choice is
quite limited by the requirement that \astrowisp{} should work on the three
major operating systems, and that it should assume as little as possible
about the images or users. The best free and open-source option we found was
the \texttt{fistar} tool of the \texttt{FITSH} package \citep{Pal_12}. Since
\texttt{FITSH} was only packaged for Linux, we incorporated the parts
necessary to compile \texttt{fistar} in \astrowisp, ensured our build system
can successfully compile the tool, and include compiled versions of the tool
in the Python wheels used to install \astrowisp.

The matching of extracted sources to an external catalog and using this match to
derive a transformation from sky to image coordinates is part of the \autowisp{}
companion package. Briefly, the current implementation begins with a crude
approximation of the transformation obtained using the \texttt{astometry.net}
package \citep{Lang_et_al_10}, either locally or the web API they provide and
then refining that transformation using it to match the extracted stars from the
input images to the third release of the Gaia catalog \citep{GaiaDR3_23}. For
the full details we refer the reader to the companion paper
\citep{Romero_et_al_25}.

Photometry starts by determining the background level and its uncertainty for
each source. \astrowisp{} begins by excluding all pixels within a certain radius
of each star and then calculating the median average and the scatter around it
after outlier rejection of pixels whose centers are within some user-specified
distance (larger than the exclusion radius). The choice of exclusion radius and
outer boundary has to be determined for the combination of telescope and camera
to avoid starlight from being included in the background determination region,
while allowing the background estimate to remain as local as possible. In all
the applications we present in Section~\ref{sec:demo} we found that an exclusion
radius of 6 pixels and an outer radius of 13 pixels works well. Those values may
need to be adjusted if stars are significantly sharper or more spread out and/or
for cameras with larger or smaller pixels.

In order to account for the effect of the Bayer mask, or of nonuniformly
sensitive pixels in general, one needs to know how the light hitting the
detector is distributed over a superpixel's surface. In astronomical images
that information is provided by a model for the PSF --- the distribution of
light produced by the optical system in the plane of the detector for a point
source at infinity.  Alternatively, one can model the PRF --- the response of a
detector pixel at some offset from the ``center'' of the source. That is, the
PRF is the PSF convolved with the sensitivity of detector pixels (for DSLR
images that means integrating the PSF over the relevant 1/4 of a superpixel).
\astrowisp{} supports both representations using a general piece-wise polynomial
model.

The area around the source location (as reported by astrometry) is split into a
rectangular grid of cells (unrelated to pixels) with, in general, nonuniform
sizes, organized in M rows and N columns. In each cell, the amount of light
falling on the detector is modeled as a bicubic function.

Formally, let $x_i$ and $y_i$ be the grid boundaries along the $x$ and $y$ image
directions respectively. Then the distribution of light from a source reported
by astrometry to be centered at $(x_0, y_0)$ for the grid cell $x_i < x - x_0 <
x_{i+1}$, $y_j < y - y_0 < y_{j+1}$ is given by:
\vspace{-4mm}\begin{equation}
    f_{i,j}(x,y)=\sum_{m=0}^3\sum_{n=0}^3 C_{i,j,m,n} x^m y^n \label{eq:psf}
\vspace{-3mm}\end{equation}
where which $i$ and $j$ is used when evaluating the PSF is determined by which
grid cell $(x,y)$ falls into. The $C_{i,j,m,n}$ coefficients specify the shape
of the PSF/PRF. Note that the grid boundaries are completely independent of
pixels. Neither source centers nor grid boundaries are assumed to line up with
pixel boundaries or pixel centers.  Instead \astrowisp{} adds up contributions
from all grid cells that overlap with a given pixel when predicting the pixel
value from the PSF/PRF model.

\astrowisp{} does not fit $C_{i,j,m,n}$ directly. Instead, some common sense
restrictions are imposed. First, $f(x,y)$ is assumed to be continuously
differentiable across cell boundaries and second, the PSF and all its
derivatives are assumed to be zero at the grid boundaries. Under these
assumptions, the PSF or PRF model is fully specified by the values of $f$,
$\frac{\partial f}{\partial x}$, $\frac{\partial f}{\partial y}$ and
$\frac{\partial^2 f}{\partial x \partial y}$ on the $(N-1)\times(M-1)$ interior
cell corners, and all these are assumed zero on the outer grid boundary. In
\astrowisp, the PSF/PRF model is split into two components:
\begin{itemize}
    \item[Shape:] the sum in Eq.~\ref{eq:psf}, constrained to have an integral
        of one.
    \item[Flux:] an overall scaling of the shape to match the observed pixel
        values of that star. This is one of the methods of measuring the
        brightness of stars in \astrowisp.
\end{itemize}

The \fitstarshape{} tool of \astrowisp{} derives the parameters of the PSF/PRF
in an image or series of images, while properly accounting for overlapping
sources. In addition, in the case of PSF fitting, \fitstarshape{} properly
handles nonuniformly sensitive pixels (or Bayer mask as a special case). PRF
fitting does not require knowledge of the sensitivity map of pixels, since that
is included in the definition of the PRF.\@ However, aperture photometry
requires separate PSF and pixel sensitivity information. \astrowisp{} uses exact
analytical expressions, ensuring machine precision calculation without incurring
a speed penalty. The parameters specifying the PSF/PRF shape are constrained to
change smoothly as a function of position within the image, as well as any user
supplied property of stars (e.g.\ color), usually taken from the external
catalog used for astrometry.  Optionally, the PSF/PRF shape can also be
constrained to change smoothly between images.

\subsection{Aperture Photometry}%
\label{sec:aperture_photometry}

One of the commonly used methods to measure fluxes of sources from an image is
to sum--up the flux within some circular aperture centered on each source
position. The reason for choosing circular apertures is their mathematical
simplicity, and that typically, point sources produce roughly circularly
symmetric profiles. In the case of wide-field images, stars are usually circular
near the center, and become significantly elongated in the outskirts (see for
example Fig. \ref{fig:color_image}) .  However, even in this case, circular
apertures are a natural choice, since the elongation direction varies with image
location, often averaging out to a circular profile over the entire image.

In general there is a trade--off that has to be made when choosing an
aperture. Large apertures include more sky noise, which affects faint
sources, and small apertures exclude flux from bright sources, thus
increasing the Poisson noise. Luckily, flux measurements using multiple
apertures can be performed, resulting in optimal photometry regardless of the
brightnesses of the sources.

The \subpixphot{} tool of \astrowisp{} takes a list of sources and information
about the PSF and background from the \fitstarshape{} tool, and outputs
measurements of the flux of each source and its formal error.  The flux ($F$)
and its Poisson error ($\delta F$) are estimated as $F=\sum_p k_p m_p - \pi
r^2B$ and $\delta F=\sqrt{\sum_p k_p^2 m_p g_p^{-1} + \pi r^2\delta B^2}$, where
the $p$ index iterates over all pixels which at least partially overlap with the
aperture, $m_p$ are the measured (after bias, dark, and flat-field corrections)
values of the pixels in ADU, $g_p$ are the gains of the pixels in electrons/ADU
(including the effects of the preapplied flat field correction), $k_p$ are
constants which account for the subpixel sensitivity map and pixels straddling
the aperture boundary, $r$ is the size of the aperture, and $B$ and $\delta B$
are estimates of the background and its error for the source. In calculating
$k_p$, \subpixphot{} properly integrates the PSF over subpixels and/or areas
inside vs.\ outside the aperture rather than assuming uniform flux distribution
over pixels or uniform pixel sensitivity.

If the subpixel sensitivity map is $S(x,y)$ ($0<x<1$, $0<y<1$), the PSF is
$f(x,y)$ ($x$ and $y$ are relative to the input position of the source center
from astrometry), and $l_p$/$b_p$ is the horizontal/vertical distance between
the left/bottom boundary of the $p$-th pixel and the source location, then the
$k_p$ constants are given by:
\vspace{-3mm} \begin{equation}
    k_p\equiv \frac{%
        \mathop{\mathlarger{\int_{pixel\cap aperture}}} \Big[
            f(l_p+x, b_p+y) + B
        \Big] dxdy
    }{%
        B+\mathop{\mathlarger{\int}}\limits_0^1 dx
        \mathop{\mathlarger{\int}}\limits_0^1 dy f(l_p+x, b_p+y)S(x,y)
    }
\label{eq:apphot_coef}
\end{equation}
where the integral in the numerator is performed over the part of the pixel that
is inside the aperture.

\subsection{Python Wrapper}%
\label{sec:python_wrapper}

All the tools described above are implemented in \texttt{C} and \texttt{C++}
enabling a host of optimizations that dramatically improve performance. These
tools are compiled into a shared library on Linux and MacOS platforms and a
dynamically loaded library (DLL) on Windows. In addition, a platform-specific
executable for the \texttt{fistar} tool from the \texttt{FITSH} package is also
compiled along with the library. 

In order to facilitate the use of \astrowisp{} both within our research group
and by others, we ensured that the user is not exposed to any of these details,
but interact with our tools through a native Python interface. This is
accomplished by Python bindings and wrapper classes.

\subsection{Packaging and Continuous Integration}%
\label{sec:packaging}

Compiling from source code can often lead to frustrating experience by users.
Fortunately, the Python packaging system includes a mechanism, known as wheels,
to distribute platform-specific precompiled versions of components developed in
languages that require compilation. Separate wheels must be created for
different versions of each operating system and for each Python version. As
new Python and OS versions come out on a regular basis, we automate the creation
and testing of wheels using \texttt{cibuildwheel} \citep{cibuldiwheel_24} and
the GitHub Actions continuous integration and continuous delivery platform
\citep{github_actions_24}.

\section{Example Photometry}%
\label{sec:demo}

In order to demonstrate that \astrowisp{} is capable of extracting
high-precision photometry from real-world citizen science images, we carried out
all the steps that an eventual photometry pipeline must include to extract
lightcurves from several datasets using both color and monochrome detectors.

\subsection{WASP-33 Observations by PANOPTES}
\begin{figure}[ht!]
    \includegraphics[width=0.5\textwidth]{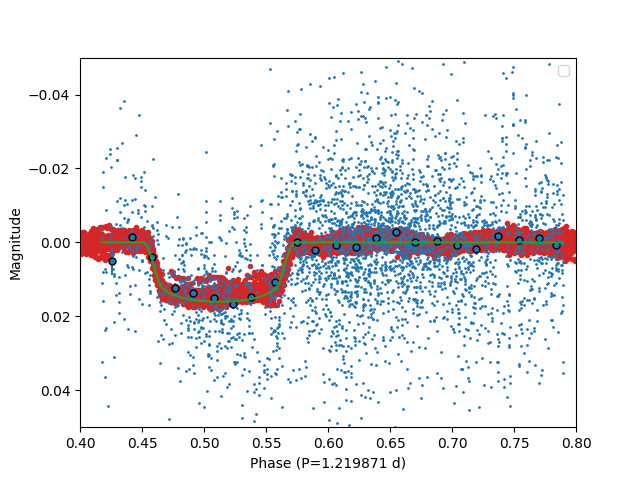}\\
    \caption{
        WASP-33 b exoplanet transit, observed by PANOPTES (blue points and
        circles), TESS (red points), and theoretical lightcurve based on best
        known system parameters (green curve). The raw PANOPTES-DSLR
        measurements, originating from the 4 color channels of four cameras in
        Hawaii (Mauna Loa observatory) and California (Mt Wilson) are shown as
        blue points. The blue points are binned in time to create the blue
        circles and corresponding error bars.
        \label{fig:wasp33}
    }
\end{figure}

The first set of images we processed were collected using over-the-counter color
DSLR cameras and lenses by the PANOPTES citizen science project. The baseline
PANOPTES model uses Canon SL2 cameras with 85mm Rokinon f/1.4 lenses, yielding a
10x15 deg field of view with 9'' sampling. The dataset consists of images
collected close to the expected transit times of the WASP-33 b exoplanet between
2018 September 11  and 2019 October 10. The project collected images not just
near transit, but we selected only those observing sessions which are expected
to include at least part of a transit. Observations from three cameras are
included, located at Mauna Loa Observatory (MLO) in Hawaii and California (Mt
Wilson) and include observations with exposure times of 30\,s, 35\,s, 60\,s, and
120\,s, totaling 1312 images.  No calibration data (dark current or flat field)
was available, so photometry was extracted from the raw images.

The resulting phase-folded lightcurve is shown in Fig.~\ref{fig:wasp33}.
Separate brightness measurements were extracted from each channel of each camera
(small blue points) which were then binned in phase to produce the larger
circles with error bars estimated as the root mean scatter around the median in
each bin.

For comparison, we also show the TESS lightcurve (red points), as well as a
theoretical model (green line) based on the literature parameters of the WASP-33
system. Note that the scatter in TESS points is not instrumental, but rather it
is intrinsic variability in the host star, which is a member of the delta-Scuti
class of variable stars.

\begin{figure}[ht!]
    \includegraphics[width=0.5\textwidth]{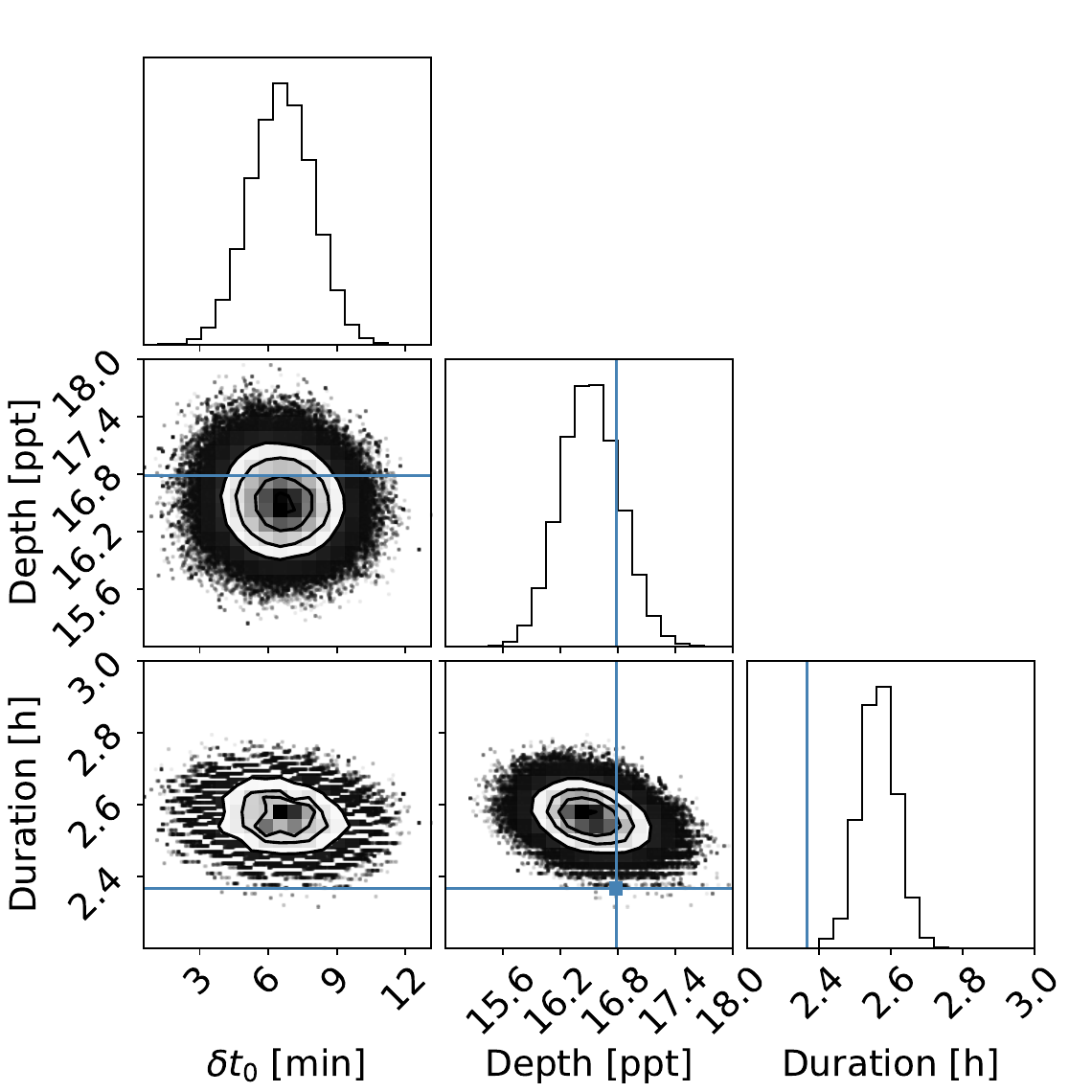}\\
    \caption{
        Posterior distribution of the transit timing, depth and duration of
        WASP-33 b using only the PANOPTES observations from Fig.
        \ref{fig:wasp33}. The timing is shown as offset from the nominal
        literature ephemeris and the blue lines show the nominal literature
        values for the depth and duration.
        \label{fig:wasp33_corner}
    }
\end{figure}

In order to quantify the achievable precision of transit observations, we
carried out a Markov Chain Monte Carlo (MCMC) analysis of the PANOPTES
lightcurve using the \texttt{emcee} \citep{Foreman-Mackey_et_al_13} and the
\texttt{pytransit} \citep{Parviainen_2015} Python packages. In order to get the
most conservative estimate of the precision, we used uniform completely
uninformative priors for the transit parameters: midtransit time,
planet-to-star radius ratio, ratio of semimajor axis to stellar radius,
inclination, and limb-darkening coefficients (assuming a quadratic
limb-darkening model). The only parameter we kept fixed was the orbital period
since for these observations that parameter is mostly degenerate with the
transit epoch.  Fig.\ref{fig:wasp33_corner} shows the posterior distribution of
the transit timing, depth, and duration. The timing is shown as an offset from
the nominal literature ephemeris: $T_{mid} = 2454163.22367 + n \times
1.21987070$ \citep{Zhang_et_al_18}, where $T_{mid}$ is the midtransit time, and
$n$ is a running index of the transits. The depth was defined as the mean of the
transit model for the central 50\% of the time between first and fourth contact
and the duration was defined as the amount of time the system spends below half
the transit depth. Based on these results, the midtransit time occurs $6.59
^{+1.37}_{-1.35}$ minutes later than the literature ephemeris, the depth is
$16.51^{+0.32}_{-0.30}$ ppt, matching very well to the literature value of 16.78
if we use our definition of the depth. The duration we find is $2.56\pm0.05$
hours. Using our definition, the literature value of 2.37 hours is slightly
shorter (by about 3.8-$\sigma$).  This slight discrepancy is likely due to the
visible tail of downward outlier points in Fig.~\ref{fig:wasp33}. We do not
attempt to quantify the ingress or egreess duration, since we do not expect that
data with the limited signal-to-noise ratio evident in Fig. \ref{fig:wasp33} to
provide meaningful constraints on these parameters.

Clearly the transit of WASP-33~b is easily detected with the correct parameters
and with sufficient precision for ephemerides maintenance to allow efficiently
scheduling follow-up observations. This is achieved even using off-the-shelf
cameral lenses as telescopes, with color detectors, without calibration data,
and not keeping the exposure time constant. Presumably significantly better
photometry can be produced with calibration and by using a fixed exposure time.

\subsection{General Monitoring Campaign by PANOPTES}
\begin{figure}
    \includegraphics[width=0.5\textwidth]{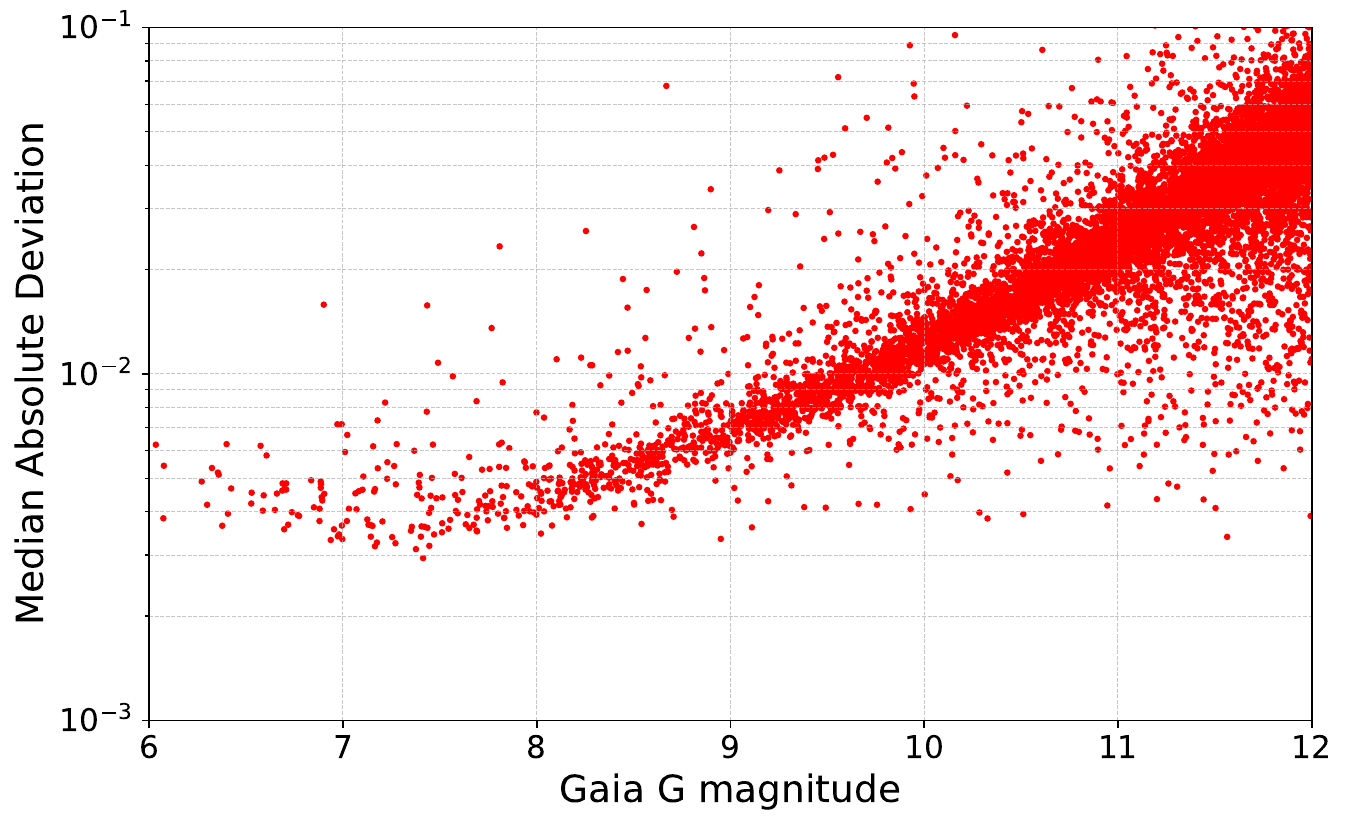}
    \includegraphics[width=0.5\textwidth]{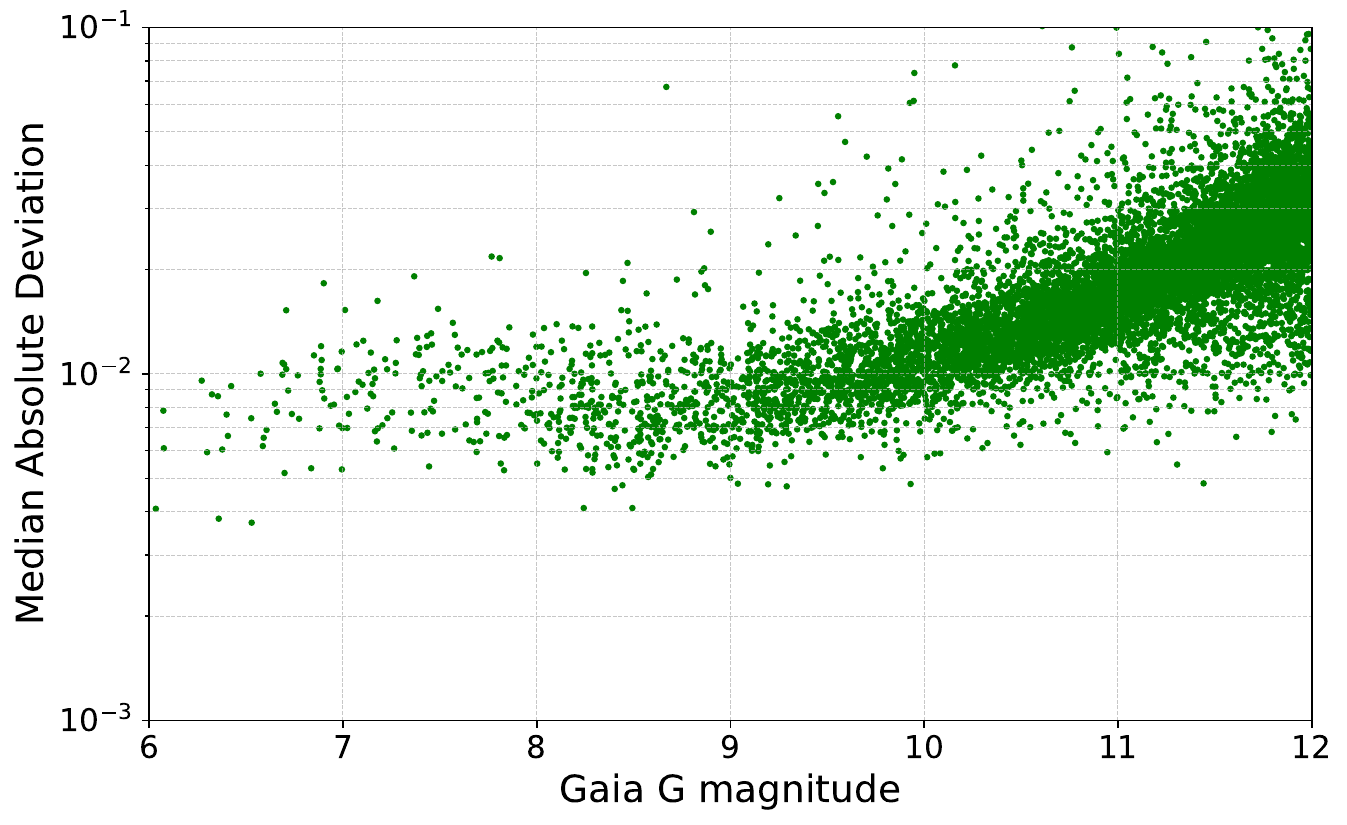}
    \includegraphics[width=0.5\textwidth]{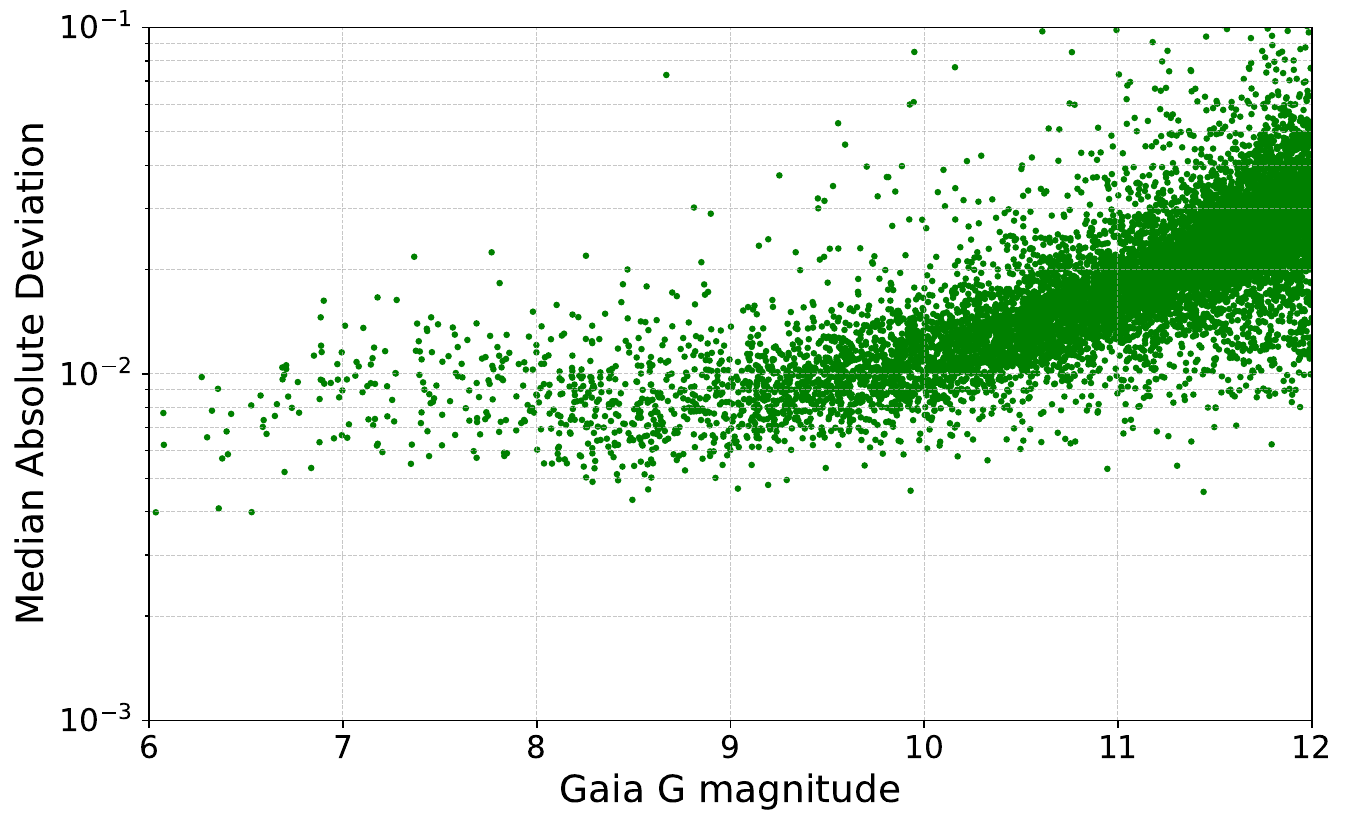}
    \includegraphics[width=0.5\textwidth]{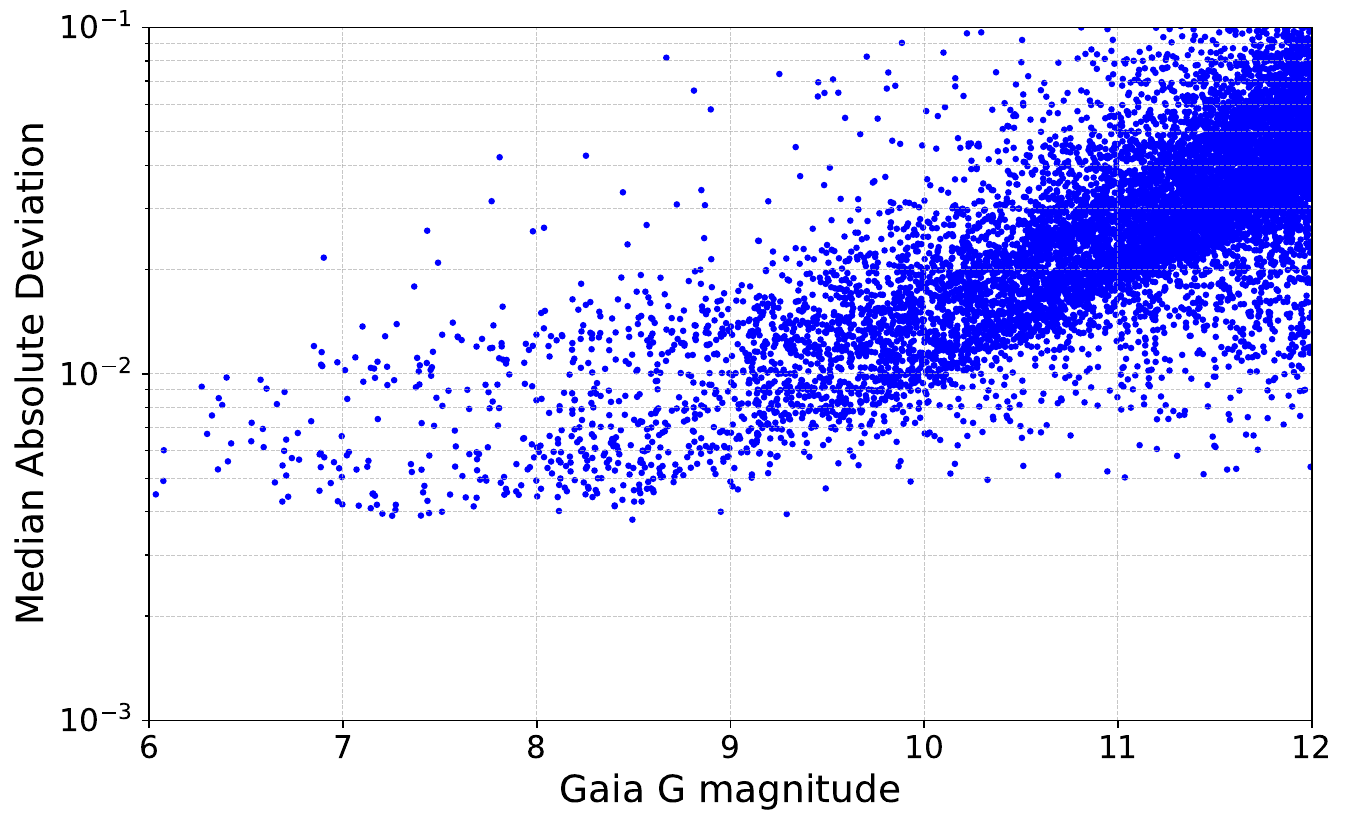}
    \caption{
        The scatter (median absolute deviation from the median) of the
        individual channel lightcurves of PANOPTES observations of a
        $10^\circ\times15^\circ$ field centered on FU Orionis. From top to
        bottom: red channel, first green channel, second green channel, blue
        channel.
        \label{fig:FUOri_channels}
    }
\end{figure}

\begin{figure}
    \includegraphics[width=0.5\textwidth]{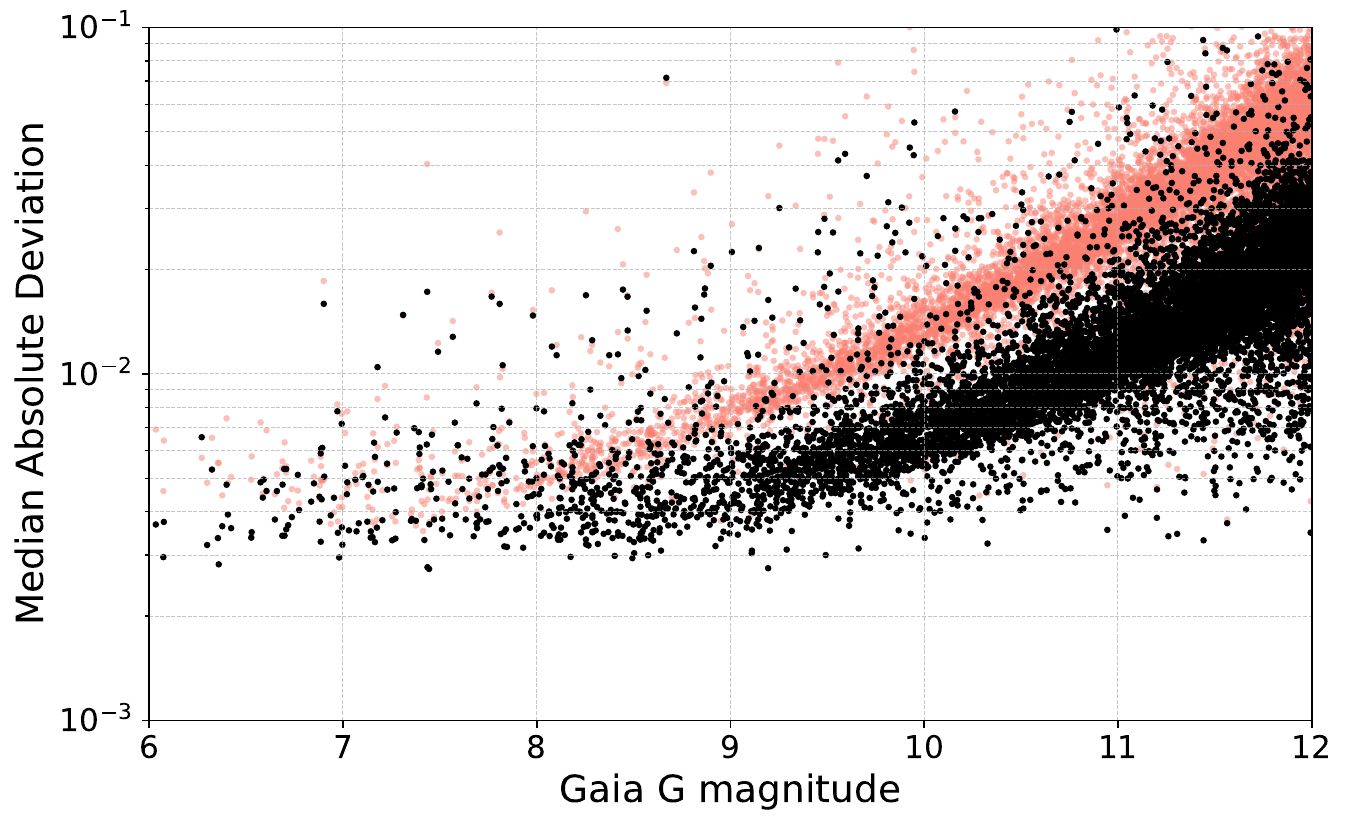}
    \caption{
        Same as Fig.~\ref{fig:FUOri_channels} but for the lightcurves combining
        the measurements in all channels (black points). For comparison the red
        channel scatter is also shown (red points). Comparison of the scatter
        between all pairs of channels as well as each channel against the
        combined scatter is available as an online figure set.
        \label{fig:FUOri_combined}
    }
\end{figure}

For an additional demonstration of the photometric precision achieved using
\astrowisp{} from color DSLR images, we processed another dataset from Project
PANOPTES, consisting of 1026 images, all with exposure time of 2 min, collected
by a single camera from a PANOPTES unit located at MLO. The data was collected
over a period of several months during 2022.  Lightcurves were created for all
stars brighter than Gaia G magnitude of 12.  Five different sets of lightcurves
were created: one for each color channel (Fig.~\ref{fig:FUOri_channels}) as
well as a combined one using the weighted average of the channels for each star
in each image (Fig.~\ref{fig:FUOri_combined}). The scatter for a given
lightcurve was estimated by taking the median of the absolute deviation from the
median of the measured brightness in each 2\,min exposure.

From the above figures we see that \astrowisp{} enables a few parts per thousand
photometric precision per exposure even from images with Bayer masks,
significantly outperforming prior efforts \citep[e.g.][]{Guyon_Martinache_12,
Zhang_et_al_15}. Even individual color channels result in better than 1\%
photometry per 2\,min exposure.

%

\section{Discussion}%
\label{sec:discussion}

This article describes a free and open-source package, called \astrowisp{},
which provides those essential components of an eventual photometry pipeline
that must be compiled for efficiency reasons. The goal is to enable high
precision photometry from color detectors like digital single-lens reflex
cameras (DSLRs), while remaining user-friendly and available for all major
operating systems.

We demonstrated that these tools, when incorporated in a full end-to-end
processing of photometric survey images, delivers high-precision lightcurves
even from a very challenging dataset.

The components described here, unlike a Python-only package, require generating
numerous wheels to support the wide range of operating systems, hardware, and
Python versions. Generating these wheels is both labor intensive and limited by
the free allotment of GitHub Actions minutes. At the same time, \astrowisp{}
represents the most stable part of the pipeline, with far fewer updates
necessary compared to the remaining code-base. For these reasons, the optimal
strategy to achieve ease of maintenance and ensure broad support across hardware
and software platforms is to distribute \astrowisp{} as a separate Python
package, rather than as part of the complete pipeline, requiring far fewer
versions to be provided and hence generating far fewer wheels by the developers,
and fewer updates to be installed by users.

\begin{acknowledgments}
    This work was supported by the National Science Foundation under grant No.
    2311527.
\end{acknowledgments}

%


\software{
    astropy \citep{astropy_13, astropy_18, astropy_22},
    fitsh \citep{Pal_12},
    astrometry.net \citep{Lang_et_al_10}
}



\bibliography{bibliography}{}
\bibliographystyle{aasjournal}



\end{document}